# Knowledge Mapping in Electricity Demand Forecasting: A Scientometric Insight

Dongchuan Yang[a], Ju-e Guo[a], Jie Li[b], Shouyang Wang[c,d,e],Shaolong Sun[a*]
[a]School of Management, Xi'an Jiaotong University, Xi'an, 710049, China
[b]College of Ocean Science and Engineering, Shanghai Maritime University, Shanghai, 201306, China
[c]Academy of Mathematics and Systems Science, Chinese Academy of Sciences, Beijing 100190, China
[d]School of Economics and Management, University of Chinese Academy of Sciences, Beijing 100190, China
[e]Center for Forecasting Science, Chinese Academy of Sciences, Beijing 100190, China
*Corresponding author. School of Management, Xi'an Jiaotong University, Xi'an 710049, China.
Tel.: +86 15911056725; fax: +86 29 82665049.
E-mail address: sunshaolong@xjtu.edu.cn (S. L. Sun).

**Abstract:** Forecasting electricity demand plays a fundamental role in the operation and planning procedures of power systems and the publications about electricity demand forecasting increasing year by year. In this paper, we use Scientometric analysis to analyze the current state and the emerging trends in the field of electricity demand forecasting form 831 publications of web of science core collection during 20 years: 1999–2018. Employing statistical description, cooperative network analysis, keyword co-occurrence analysis, co-citation analysis, cluster analysis, and emerging trend analysis techniques, this article gives the most critical countries, institutions, journals, authors and publications in this field, cooperative networks relationships, research hotspots and emerging trends. The results of this article can provide meaningful guidance and some insights for researchers to find out crucial research, emerging trends and new developments in this area.

## 1. Introduction

Nowadays, electricity has become one of the most important energy and plays an indispensable role in many fields. In recent years, a large number of researches have proved that the accuracy of electricity demand prediction plays a fundamental role in the operation and planning procedures of power systems[1, 2]. Accurate electricity demand forecasting can not only ensure the reliable operation of power systems but also have a great cost-saving potential for power corporations [3]. On a temporal scale, electricity demand forecasting can be classified into short-term, medium-term and long-term. With the increase of electricity demand and the rapid development of artificial intelligence technology, electricity demand forecasting has attracted more and more scholars' attention and new research methods, emerging trends and new developments have been generated at the same time [4]. A large number of forecasting researches and methods have been proposed and applied in the field of electricity load forecasting in recent years [5, 6].

Mohan et al. [7] proposed a data-driven method for short-term load forecasting using dynamic mode decomposition. Musaylh et al. [8] proposed a hybrid model that including the multivariate adaptive regression, artificial neural network and multiple linear regression models to forecast short-term electricity demand in Australia. Shao et al. [9] conducted decomposition methods for electricity demand forecasting and presented that Empirical mode decomposition and wavelet decomposition are the most popular technique. Kuster et al. [1] presented a review which revealed that artificial neural network, multivariate regression, time series analysis and multiple linear regression are popular and effective methods for electricity and electricity forecasting. Hong and Fan [10] offered a tutorial review of probabilistic electric load forecasting and introduced the techniques, methodologies, applications, evaluation methods and future research needs. However, very little research has analyzed the collaborative relationship, new developments and  emerging trends of electricity demand forecasting and visualized the knowledge map in this field.

Scientometrics is an important method to find out the rules of scientific activities, identify research trends, and evaluate the development of the field [11, 12]. Yu and Xu analyzed the current status and explore future research trends of the carbon emission trading domain by the scientometric method [13]. Olawumi and Chand evaluated the research development status of institutions, countries, and journals in the research field [12]. Niazi and Hussain evaluated all sub-domains of agent-based computing and found agent-based computing extensive in other dominos [14].

With the rapid growth of attentions and publications for electricity demand forecasting, it is necessary and urgent to summarize the current situation and analyze the collaborative relationship, new developments and  emerging trends of electricity demand forecasting. In this paper, scientometrics analysis is performed in the electricity load forecasting domain and utilizing software named CiteSpace to analyze and visualization the emerging trends. CiteSpace, invented by Dr. Chen Chaomei, is a particularly popular method of scientometrics that uses citation analysis in a visual form and can be used to identify knowledge areas and emerging trends in researches [15, 16]. In recent years, CiteSpace has attracted the interest of many scholars and has been applied to many fields. Chen used published literature to investigate Emerging trends and new developments in regenerative medicine [17]. Yang et al. comprehensively analyzed the status of PM2.5 research and found the frontiers of research in this field [18]. Fang et al. analyze the interaction between climate change and tourism and describe the research characteristics of the field in the past 25 years [19].

The structure of this article is as follows: **Section 2** gives the source and search strategy of publications. **Section 3** introduces the basic summary of electricity demand forecasting research. In **Section 4**, we visualize the cooperation network of authors, institutions, and countries. **Section 5** analyzes the active topics and emerging trends in electricity load forecasting, including keyword analysis and co-citation analysis. **Section 6** gives comprehensive conclusions and discussions.

## 2. Methodology

This section provides the search strategy of data. For the searched phrase in WOS, some articles perform an exact search on a certain phrase, such as Yu and Xu, and some articles perform an exact search on multiple phrases and merge the results, such as Chen [13, 20]. Few articles obtain articles through searching inexact themes, because searching inexact themes requires that the query words do not have to appear consecutively, which gets a large number of publications that are not related to the search subject. It is worth noting that this article searches precise themes and non-precise titles. This article focuses on a more subdivided field, and the number of related articles is little. Searching precise themes will ignore indispensable publications in the field of electricity load forecasting and affect the conclusion of this article seriously. To improve the recall rate and avoid a large number of irrelevant publications being retrieved, this article adopts the strategies of searching precise themes and inexact titles which ensures the accuracy of publications being retrieved through means of manual screening.

The data used for analysis in our research is downloaded from Web of Science (WoS), and the search strategy we followed is below:

(1) (TS=("electricity demand forecasting" OR “electricity demand prediction" OR "electrical demand forecasting" OR "electrical demand prediction" OR "electric demand forecasting" OR "electric demand prediction" OR "power demand forecasting" OR "power demand prediction" OR "electricity consumption forecasting" OR "electricity consumption prediction" OR "electrical consumption forecasting" OR "electrical consumption prediction" OR "electric consumption forecasting" OR "electric consumption prediction" OR "power consumption forecasting" OR "power consumption prediction" OR "electricity load forecasting" OR "electricity load prediction" OR "electrical load forecasting" OR "electrical load prediction" OR “electric load forecasting" OR "electric load prediction" OR "Electricity load forecasting" OR "Power load prediction" OR "grids load forecasting" OR "grids load prediction") OR TI=(electricity demand forecasting OR electricity demand prediction OR electrical demand forecasting OR electrical demand prediction OR electric demand forecasting OR electric demand prediction OR power demand forecasting OR power demand prediction OR electricity consumption forecasting OR electricity consumption prediction OR electrical consumption forecasting OR electrical consumption prediction OR electric consumption forecasting OR electric consumption prediction OR power consumption forecasting OR power consumption prediction OR electricity load forecasting OR electricity load prediction OR electrical load forecasting OR electrical load prediction OR electric load forecasting OR electric load prediction OR Electricity load forecasting OR Power load prediction OR grids load forecasting OR grids load prediction) NOT TS=(WLAN OR CPU OR GPU OR vehicle));

(2) Databases=Science Citation Index Expanded or (SCI-EXPANDED), Social Sciences Citation Index (SSCI), Conference Proceedings Citation Index-Science (CPCI-S) and Conference Proceedings Citation Index-Social Science & Humanities (CPCI-SSH);

(3) Timespan = "1999-2018";

(4) Document types = "article" or "review";

(5) Literature type = "English";

901 publications are retrieved and 70 publications that were not related to electricity demand forecasting were deleted through means of manual screening. Finally, 831 publications were downloaded on October 18, 2019.

## 3. Basic summary of electricity demand forecasting research.

This section provides the statistical analysis form five parts including distribution of time, subject categories, high-yield journals, high-yield institutions, high-yield authors and highly cited publications in electricity demand forecasting.

### 3.1 The distribution of publications

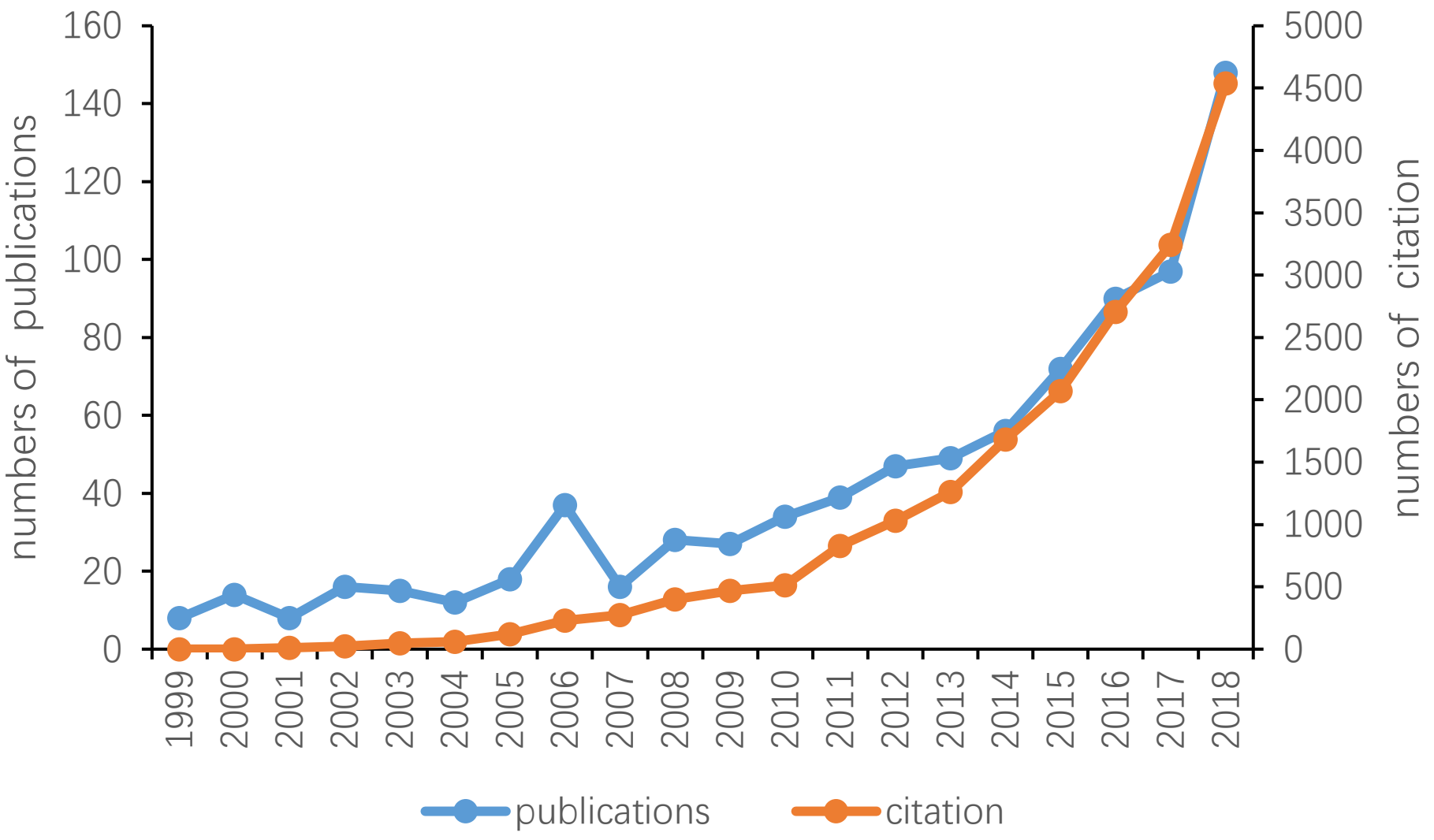

Fig.1. Numbers of publications and citations in electricity demand forecasting, 1999-2018.

The temporal distribution of publications and citations in electricity demand forecasting is shown in Fig.1. The number of publications is increasing over the past 20 years, from 8 in 1999 to 148 in 2018, with steady growth in 199-2009 and rapid growth in 2010-2018. The publications have been cited 19506 times from 1999 to 2018. The number of citations is increasing, year by year, and has similar growth trends with the numbers of publications. From this, we can see that the field has received more and more attention, especially in the last decade.

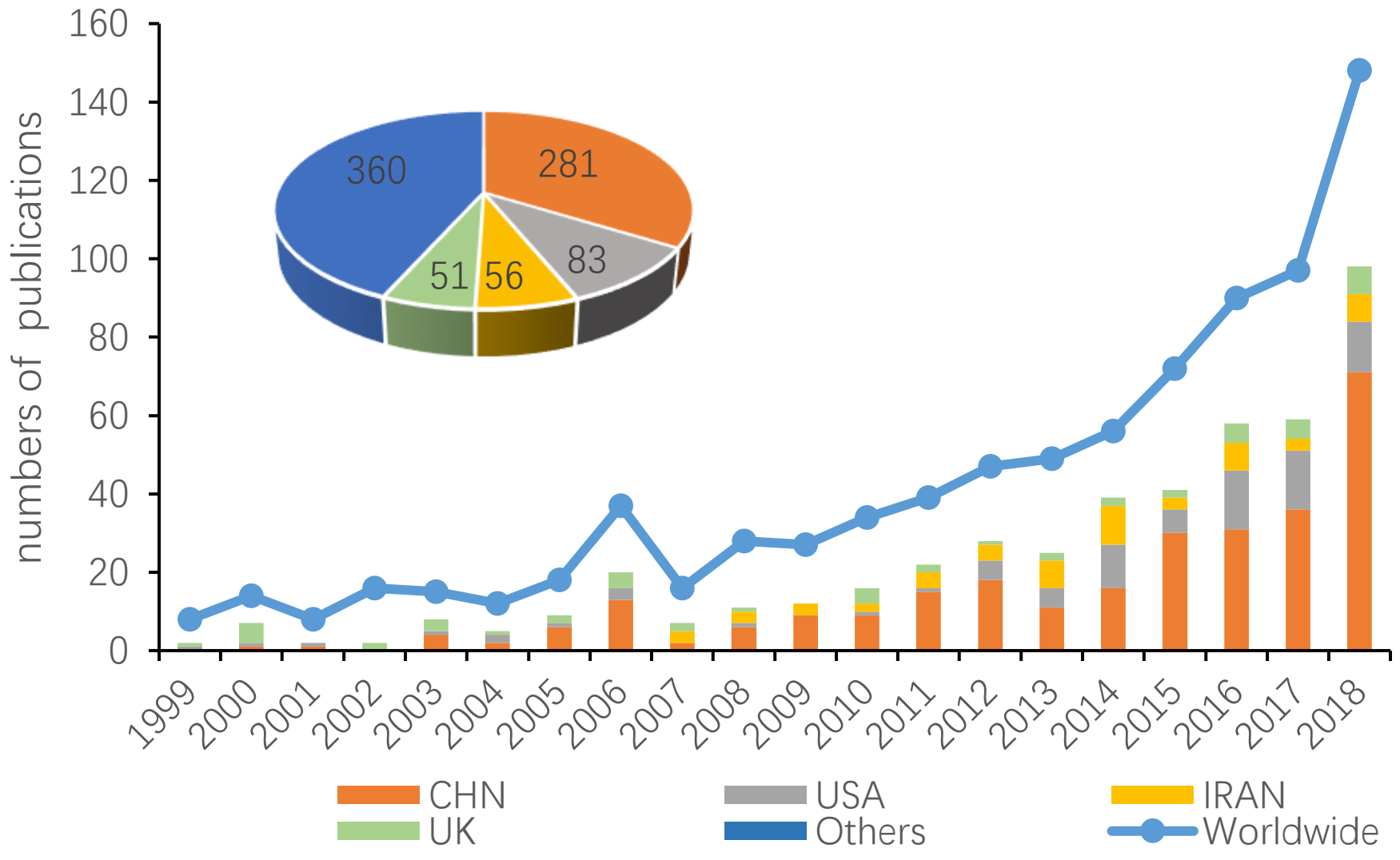

Fig.2. Distribution of publications in electricity demand forecasting, 1999-2018.

Fig.2 shows that China, the USA, Iran and the UK are the main countries publishing papers in this field and increase rapidly, similar to publications. China is the country with the most publications, especially after 2015, the number of publications in China exceeds the sum of the United Kingdom, the United States and Iran. It should be noted that the publications of Taiwan and Hong Kong are included in China. The number of publications in the United States and the United Kingdom is fluctuating. Iran has published first publication in 2007, and Iran has published more than three papers each year. In 1999-2018, China published 33.81%(281) of the total publications in electricity demand forecasting, the US for 9.99%(83), Iran for 6.74%(56) and the UK for 6.14%(51).

### 3.2 Subject categories

Fig.3 shows clearly that electricity demand forecasting is a cross-disciplinary research area, including energy fuels accounting for 36.82%(306), engineering electrical electric accounting for 26.23%(218), computer science artificial intelligence accounting for 16.49%(137), thermodynamics accounting for 13.48%(112) and economics accounting for 6.38%(53).

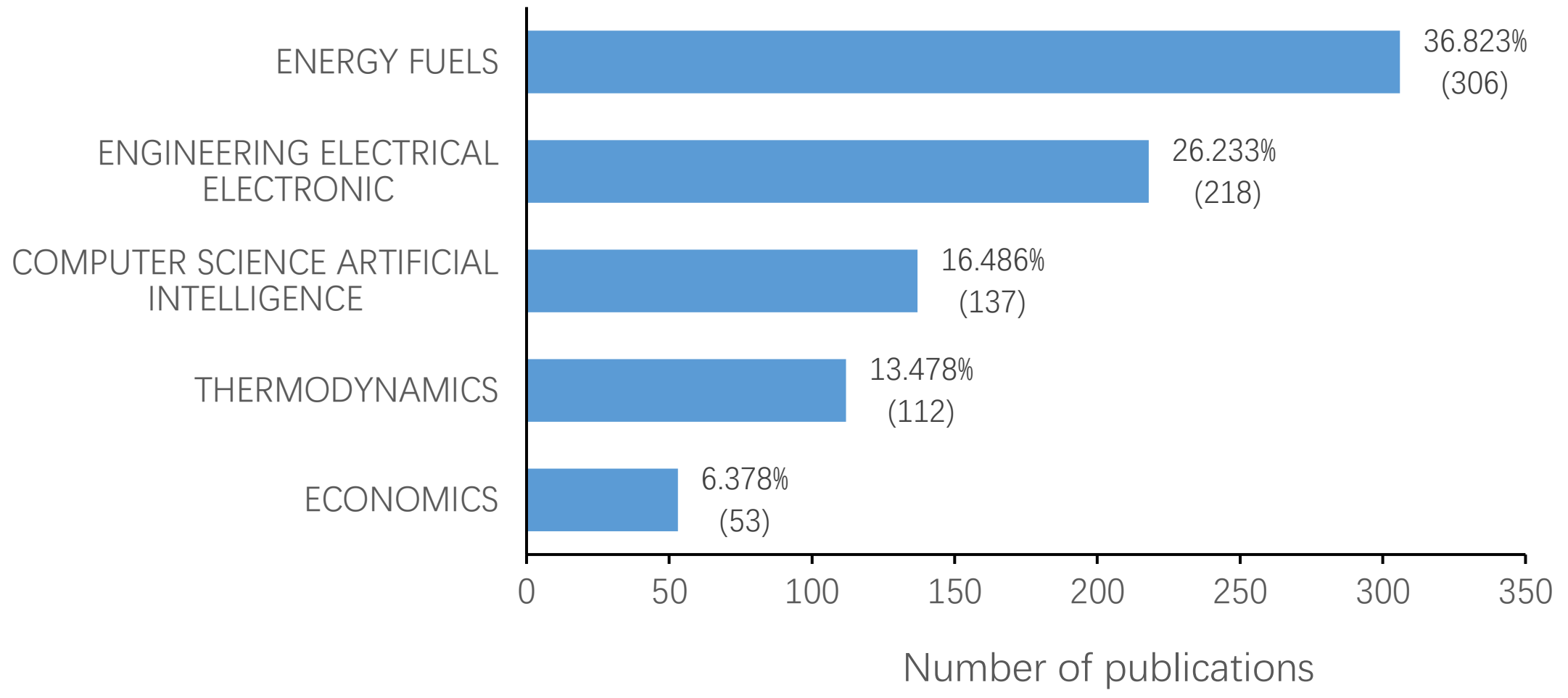

Fig.3. Distribution of the main Subjects in electricity demand forecasting, 1999-2018.

### 3.3 High-yield journals

199 journals published papers in electricity demand forecasting form 1999 to 2018. Table 1 lists the top 10 journals, it can be seen that energy and power are areas of greatest concern to the top 10 journals. “Energy” is the highest yield journal with 81 publications, followed by “Energies”, “International Journal of Electrical Power Energy Systems”, “Applied Energy”, “Energy Conversion and Management”,” Electric Power Systems Research”,” Energy and Buildings”,” International Journal of Forecasting”,” IEEE Transactions on Power Systems” and ” Lecture Notes in Computer Science”. In the top 10 journals, the impact factor of “Energy”,” Applied Energy”,” Energy Conversion and Management” and ” IEEE Transactions on Power Systems” are all more than 5.

Table 1. High-yield journals in electricity demand forecasting.

| Num | Journal | Numbers of publications | Proportion | Impact factor | Country |
|---|---|---|---|---|---|
| 1 | Energy | 81 | 9.75% | 5.537 | England |
| 2 | Energies | 63 | 7.58% | 2.707 | Switzerland |
| 3 | International Journal of Electrical Power Energy Systems | 46 | 5.54% | 4.418 | England |
| 4 | Applied Energy | 39 | 4.69% | 8.426 | England |
| 5 | Energy Conversion And Management | 29 | 3.49% | 7.181 | England |
| 6 | Electric Power Systems Research | 26 | 3.13% | 3.022 | Switzerland |
| 7 | Energy and Buildings | 23 | 2.77% | 4.495 | Switzerland |
| 8 | International Journal of Forecasting | 23 | 2.77% | 3.386 | Netherlands |
| 9 | IEEE Transactions on Power Systems | 22 | 2.65% | 6.807 | USA |
| 10 | Lecture Notes in Computer Science | 19 | 2.29% | 0.402 | USA |

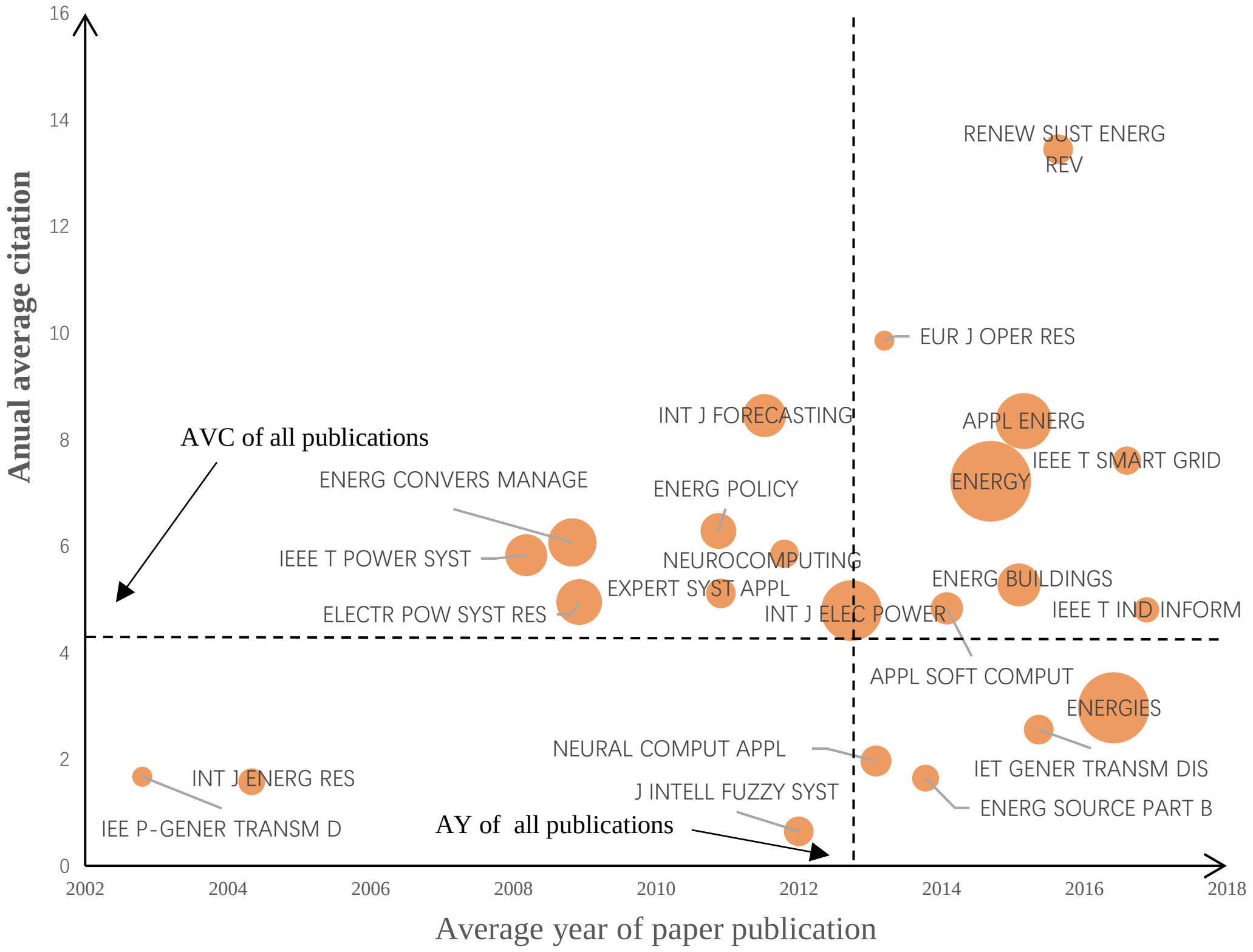

Fig.4. Distribution of main journals in electricity demand forecasting, 1999-2018.

Fig.4 shows the distribution of main journals in electricity demand forecasting. There are 29 journals in Fig.4 and each of them published more than or equal to 5 publications. We denoted with NP the number of publications, T the publication time for a publication, NC the number of citations for a publication from 1999 to 2018. Average time for a journal, $AY = \sum T / NP$, represents the horizontal axis in Fig.4. Annual average citation for a journal, $AVC = \left( \sum \frac{NC}{(2019-T)} \right) / NP$, represents the vertical axis in Fig.4. The intersection of the horizontal and vertical axes is (2012.76, 4.18). The average publication time of 831 articles is 2012.76. The average number of citations for 831 articles is 4.18. The journal above the horizontal axis indicates that the average number of citations for the journal is higher than the average in this field. The journal on the right side of the vertical axis indicates that the average publication time for the journal is closer. The size of the point in Fig.4 represents the number of publications for a journal form 1999-2018. Journals in the first quadrant, “Renewable & Sustainable Energy Reviews”, “European Journal of Operational Research”, “Applied Energy”, “IEEE Transactions on Smart Grid, Energy”, “Energy and Buildings”, “Applied Soft Computing”,” IEEE Transactions on Industrial Informatics” are of constant interest to researchers in the field. It is obvious that the annual average citation for “Renewable & Sustainable Energy Reviews” (13.45) is significantly higher and “Energy” and “Applied Energy” have more publications.

From Figure 4, we can find out which journals in the field of power forecasting have published highly cited publications recently, which journals have published highly cited publications in the past, which journals have published a large number of articles, and which journals we do not need to pay attention to.

### 3.4 High-yield institutions

848 institutions published papers in electricity demand forecasting form 1999 to 2018. Table 2 lists the top 10 institutions, it can be seen that 5 of the top 10 institutions come from China and China published the largest number of articles, which is also the same conclusion as figure 2. North China Electric Power University is the highest yield Institution with 57 publications, followed by Lanzhou University, University of North Carolina, Islamic Azad University, Oriental Inst Technol, Electricite de France edf, University of Tehran, University of Oxford, Dongbei University of Finance Economics, Hefei University of Technology.

Table 2. High-yield institutions in electricity demand forecasting.

| INSTITUTION | Country | Number of articles | Share within the country /% | Share in the world /% |
|---|---|---|---|---|
| North China Electric Power University | CHINA | 57 | 20.28% | 6.86% |
| Lanzhou University | CHINA | 35 | 12.46% | 4.21% |
| University of North Carolina | USA | 24 | 29.27% | 2.89% |
| Islamic Azad University | IRAN | 18 | 32.14% | 2.17% |
| Oriental Inst Technol | CHINA | 15 | 5.34% | 1.81% |
| Electricite De France Edf | FRANCE | 14 | 56.00% | 1.69% |
| University of Tehran | IRAN | 14 | 25.00% | 1.69% |
| University of Oxford | UK | 13 | 25.49% | 1.56% |
| Dongbei University of Finance Economics | CHINA | 13 | 4.63% | 1.56% |
| Hefei University of Technology | CHINA | 13 | 4.63% | 1.56% |

### 3.5 High-yield Authors

Table 3 shows high-volume authors, mainly from China, the United States and Iran. The publication status of the author in this field is shown in the table, where * represents the H index of the author in this field (831 pieces of literature).

Table 3. High-yield authors in electricity demand forecasting.

| Author | Country | Total number of published literatures | Total number of citations | Mean citations per article | Maximum number of citations | H index | H index** |
|---|---|---|---|---|---|---|---|
| Wang JZ | China | 24 | 864 | 35.21 | 95 | 42 | 16 |
| Hong WC | China | 21 | 1473 | 70.14 | 254 | 31 | 17 |
| Niu DX | China | 16 | 370 | 23.13 | 181 | 18 | 10 |
| Hong T | USA | 11 | 611 | 55.55 | 180 | 15 | 11 |
| Azadeh A | Iran | 11 | 490 | 44.55 | 141 | 32 | 8 |
| Amjady N | Iran | 10 | 604 | 60.4 | 181 | 38 | 9 |
| Taylor JW | England | 10 | 1334 | 133.4 | 269 | 23 | 9 |
| Yang SL | China | 10 | 123 | 12.3 | 36 | 27 | 6 |
| Goude Y | France | 8 | 275 | 34.38 | 75 | 7 | 6 |
| Che JX | China | 7 | 199 | 28.43 | 71 | 10 | 6 |

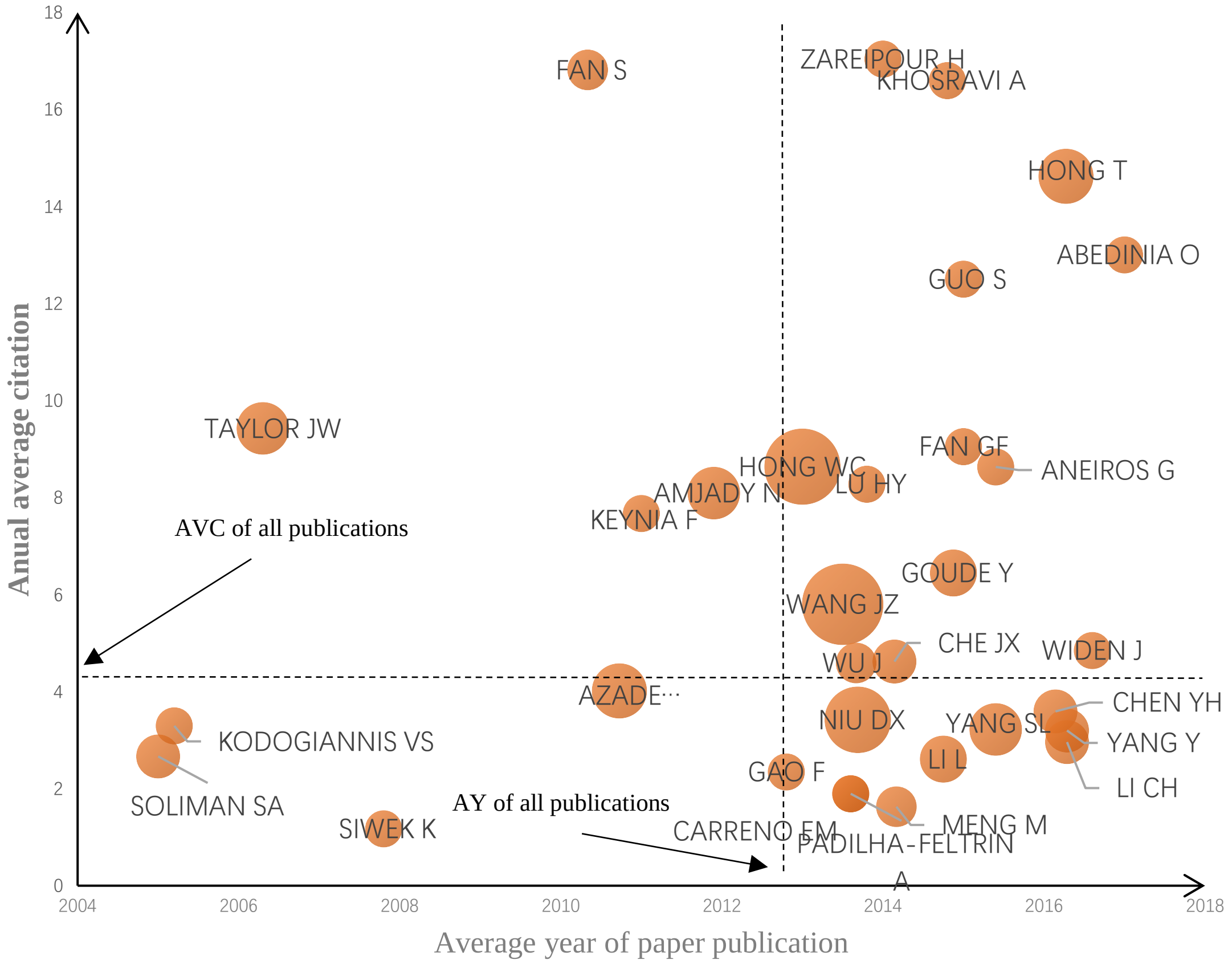

Fig.5. Distribution of main authors in electricity demand forecasting, 1999-2018.

Fig.5 shows the distribution of main authors in electricity demand forecasting. There are 32 authors in Fig.5 and each of them published greater than or equal to 5 publications. The average time for an author, $AY=\sum T/NP$, represents the horizontal axis in Fig.5. Annual average citation for an author, $AVC=\left(\sum \frac{NC}{(2019-T)}\right)/NP$, represents the vertical axis in Fig.5. The intersection of the horizontal and vertical axes is (2012.76, 4.18). The average publication time of 831 articles is 2012.76. The average number of citations for 831 articles is 4.18. The journal above the horizontal axis indicates that the average number of citations for the journal is higher than the average in this field. The journal on the right side of the vertical axis indicates that the average publication time for the journal is closer. The size of the point in Fig.5 represents the number of publications for a journal form 1999-2018.

Fig 5 shows us the number of posts, posting time and citations of the main authors in the field. The authors in the 1, 2, and 4 quadrants are worthy of our attention, especially the authors in the first quadrant, whose articles have received a lot of attention in recent years (such as ZAREIPOURH, KHOSRAVIA, HONG T, ABEDINIAO, GUO S). The author of the second quadrant is likely to publish a much-watched article by 2012. Authors of the fourth quadrant recently published articles with low attention, but their articles may become hot spots in the future. There are also some authors (such as HONG WC, WANG JZ, HONG T) who have published a large number of articles in the field.

**3.6 Highly cited Publications**

Table 4 shows the top 10 articles with the highest citations. Obviously, only the articles published by Hahn, Heiko et al. on "EUROPEAN JOURNAL OF OPERATIONAL RESEARCH" after 2009, other high-cited articles were published before 2009 [21]. Among them, Taylor, JW, Hong, WC and other authors are shown in Figure 5. It is worth noting that Alfares, HK; Nazeeruddin, M and Taylor, JW and others paid little attention to this field after publishing articles in the field before 2009.

Table 4. Highly cited publications in electricity demand forecasting.

| Authors | Published year | Journal | Total number of citations | Average citations per year |
|---|---|---|---|---|
| Alfares, HK; Nazeeruddin, M[4] | 2002 | International Journal of Systems Science | 272 | 15.11 |
| Taylor, JW[22] | 2003 | journal of the Operational Research Society | 265 | 15.59 |
| Pai, PF; Hong, WC[23] | 2005 | Electric Power Systems Research | 239 | 15.93 |
| Bunn, DW[24] | 2000 | Proceedings of the Ieee | 236 | 11.8 |
| Hahn, Heiko; Meyer-Nieberg, Silja; Pickl, Stefan[21] | 2009 | European Journal of Operational Research | 233 | 21.18 |
| Taylor, JW; Buizza, R[25] | 2002 | Ieee Transactions on Power Systems | 231 | 12.83 |
| Akay, Diyar; Atak, Mehmet[26] | 2007 | Energy | 223 | 17.15 |
| Hsu, CC; Chen, CY[27] | 2003 | Energy Conversion and Management | 220 | 12.94 |
| Taylor, JW; de Menezes, LM; McSharry, PE[28] | 2006 | International Journal of Forecasting | 219 | 15.64 |
| Li, Hong-ze; Guo, Sen; Li, Chun-jie; Sun, Jing-qi[29] | 2013 | Knowledge-Based Systems | 218 | 31.14 |

Table 5 shows the top 10 papers with the highest average citations per year. Obviously, only the articles published by Li, Hongze et al. on "Knowledge-Based Systems" before 2013, others were published after 2013 [29]. Among them, Hong, Tao, Guo, Sen, Fan, Shu and other authors are shown in Figure 5. It is worth noting that the average annual highly cited articles are mainly concentrated in the past 5 years, indicating that the electricity load forecasting may have received more attention in the near future, or new developments have appeared.

Table 5. Highly average annual cited publications in electricity demand forecasting.

| Authors | Published year | Journal | Total number of citations | Average citations per year |
|---|---|---|---|---|
| Hong, Tao; Fan, Shu[30] | 2016 | International Journal of Forecasting | 159 | 39.75 |
| Hong, Tao; Pinson, Pierre; Fan, Shu[10] | 2016 | International Journal of Forecasting | 143 | 35.75 |
| Ahmad, A. S.; Hassan, M. Y.; Abdullah, M. P.[31] | 2014 | Renewable & Sustainable Energy Reviews | 207 | 34.5 |
| Raza, Muhammad Qamar; Khosravi, Abbas[2] | 2015 | Renewable & Sustainable Energy Reviews | 157 | 31.4 |
| Li, Hongze; Guo, Sen; Li, Chun-jie[29] | 2013 | Knowledge-Based Systems | 218 | 31.14 |
| Quan, Hao; Srinivasan, Dipti; Khosravi, Abbas[32] | 2014 | IEEE Transactions on Neural Networks and Learning Systems | 172 | 28.67 |
| Mohammadi, Mohsen; Talebpour, Faraz; Safaee, Esmaeil[33] | 2018 | Neural Processing Letters | 48 | 24 |
| Kaboli, S. Hr Aghay; Fallahpour, A.; Selvaraj, J.[34] | 2017 | Energy | 68 | 22.67 |
| Boroojeni, Kianoosh G.; Amini, M. Hadi; Bahrami, Shahab[35] | 2017 | Electric Power Systems Research | 66 | 22 |
| Hahn, Heiko; Meyer-Nieberg, Silja; Pickl, Stefan[21] | 2009 | European Journal of Operational Research | 233 | 21.18 |

## 4. Cooperative structure in the field of power demand forecasting

### 4.1 Author cooperation network

The author's cooperative network shows the cooperation of all authors in the 831 papers in the field of power demand forecasting. The node size indicates the author's published volume, and the connection between nodes represents the author's cooperative relationship. The thickness of the connection indicates the strength of the cooperation between the authors. The color of the connection between nodes and nodes corresponds to the time axis. In the figure, there are 2143 points and 3561 edges (LRF=-1.0, LBY=10, e=1.0). Obviously, there are many scholars involved in the field of power demand forecasting, but most of them only cooperate in a small scope. Many small independent networks of cooperation have not formed extensive cooperation. There is also a large independent cooperation network in the cooperative network. We extracted the four largest cooperative networks from the figure, as shown in Figure 6.

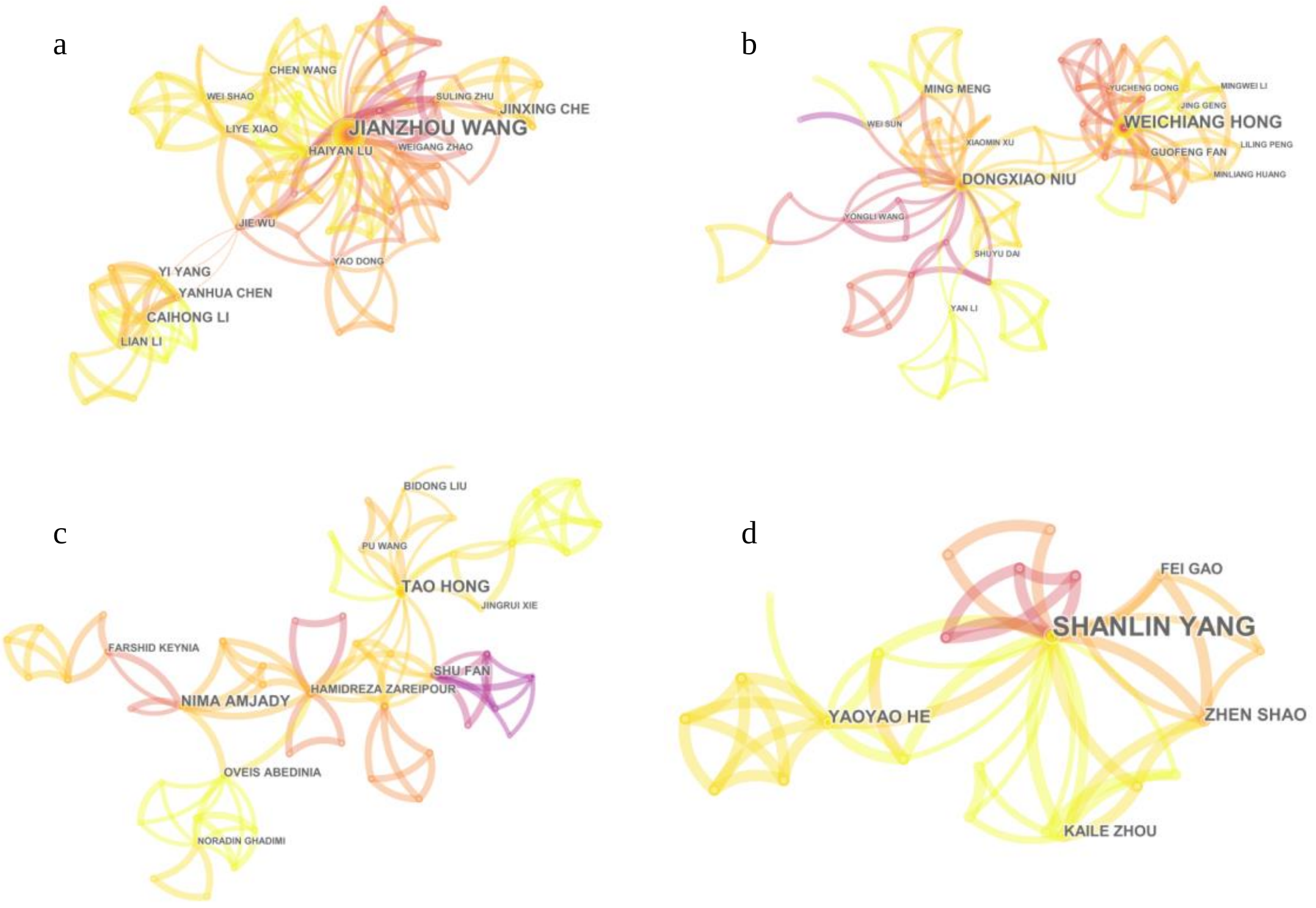

**Fig.6.** Author cooperation network in electricity demand forecasting.

**Fig.6** shows the four largest cooperative networks, with 74, 70, 53, and 24 nodes in each of the four networks. Figure a is the largest cooperative network with 74 nodes. Among them, Wang Jianzhou, Li Caihong, Yang, Y, Li, L, Wang Chen, Weigang Zhao, Wu Jie and other authors constitute a research cluster based on Lanzhou University. Figure b is the second network with 70 nodes. Among them, Dongxiao Niu, Ming Meng and other authors constitute the research cluster based on North China Electric Power University, Weichiang Hong, Fan, Guo-Feng, Peng, Li-Ling and other authors. It constitutes a research cluster based on the Pingdingshan Normal University of Jiangsu Normal University. Part c of **Fig.6** is the third network with 53 nodes. Among them, Tao Hong, Bidong Liu, Pu Wang, Jingrui Xie and other authors constitute the research cluster based on Univ of South Carolina, Farshid Keynia, Nima Amjady, Oveis Abedinia and other authors. It constitutes a research cluster based on Semnan Univ, and authors such as Shu Fan and Hamidreza Zareipour constitute a research cluster based on Univ Calgary. Authors such as Fei Gao, Shanlin Yang, Yaoyao He, Zhen Shao, and Kaile Zhou constitute a research cluster based on Hefei University of Science and Technology. Other scholars with a large number of publications and extensive cooperation include Azadeh, A of Tehran University; Goude, Y of Elect France; Taylor, JW of Oxford University. Although there are many participants, there are more networks of less than 10 partners in the cooperation network, indicating that the cooperation in the field of power demand forecasting is mostly based on small research teams.

### 4.2 Institutional cooperation network

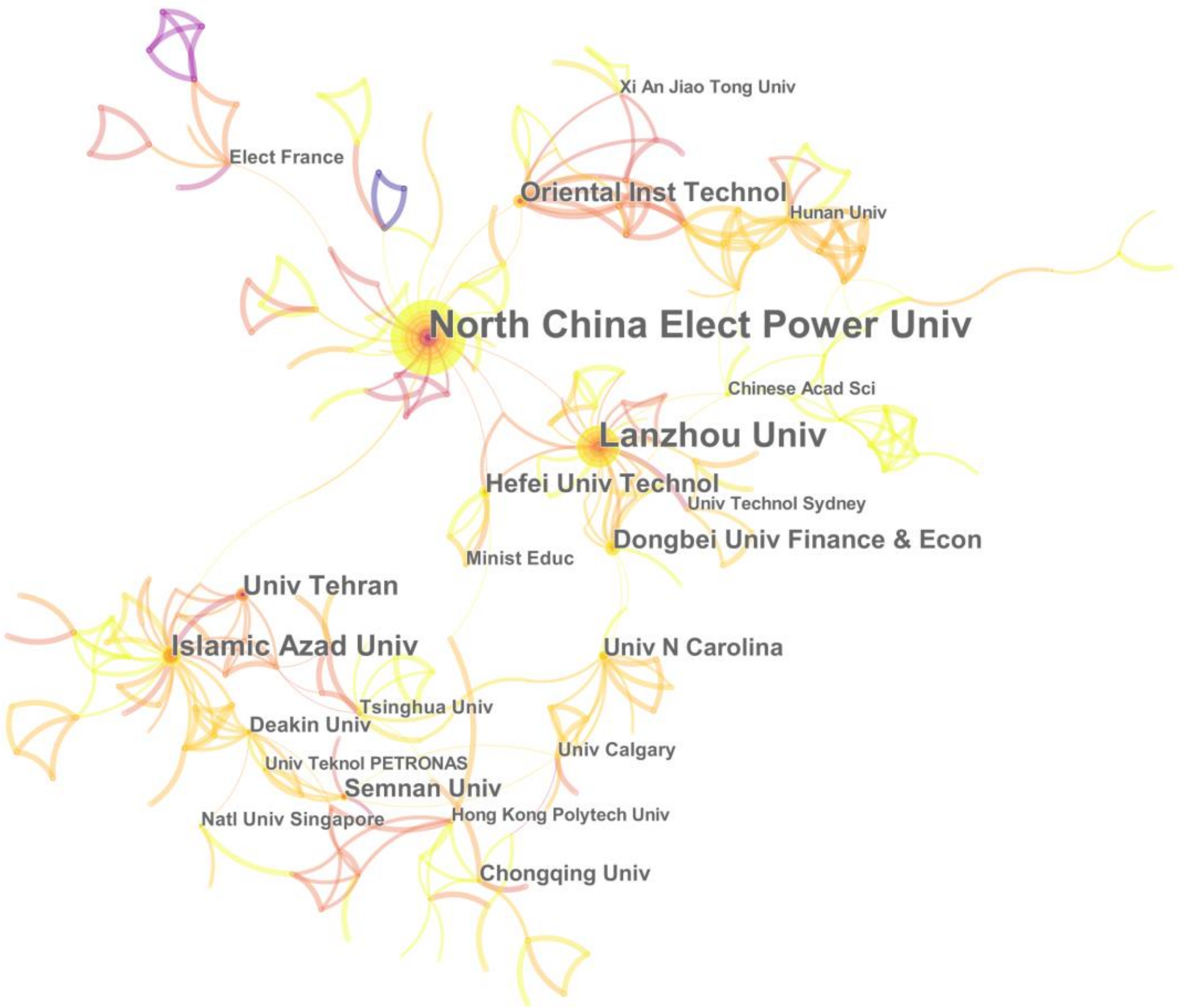

Fig.7. Institutional cooperation network in electricity demand forecasting

The power demand forecasting Institutional cooperative network has 859 points and 894 edges. Fig.7 shows the largest independent network in the institutional cooperation network, with 217 points. North China Electric Power University and Lanzhou University are the leading contributors to the cooperation in the field of power demand forecasting and the number of articles published in other aspects. After 10 connections were used as thresholds, 10 research institutions were listed: North China Electric Power University (57), Lanzhou University (35), Islamic Azad Univ (18), Oriental Inst Technol (15), Univ Tehran (14), Hefei Univ Tec2hnol (13), Dongbei Univ Finance & Econ (13), Univ Oxford (12), Semnan Univ (11), Univ N Carolina (10). From the national point of view, the main institutions in this field are more located in China and the cooperation between domestic institutions and multinational cooperation. The cooperation of Lanzhou University in China is mainly the University of Chinese Academy of Sciences, Hefei University of Science and Technology, Dongbei University of Finance and Economics. It is worth noting that Wang jianzhou is a highly productive author in this field. He has worked at Lanzhou University and Dongbei University of Finance and Economics, which also indicates the cooperative relationship between Lanzhou University and Dongbei University of Finance and Economics. Univ Tehran and Islamic Azad Univ are the main partners of each other, they are all universities in Iran. Besides, cross-border cooperation is also very common, cooperation between Lanzhou University and the University of Technology, Tsinghua University and Semnan Univ. At the same time, it is obvious that comparing the author's cooperation network, we find that the institutional cooperation network is closer.

4.3 **Country/region cooperation network**

The country's cooperative network map for electricity demand forecasting, with 37 points and 89 edges in the network (deleting links with fewer than two links), and Figure 8 shows that the largest independent network contains 30 points. The top 10 countries are Peoples R China (including Taiwan, Hong Kong and Macau), USA, Iran, England, Turkey, Spain, Australia, Brazil, Canada. From the Fig.8, we can find that China is the largest contributor to the largest network of cooperation in this field. His main partners are also countries with large volumes such as the United States (16), Australia (8) and Canada (8), the United Kingdom ( 7), Japan (4). The main partners in the United States are China (15), Italy (3), Pakistan (3), and Poland (3). The main partners in Iran are Australia (3), Malaysia (3), Canada (2), United Kingdom (2), and Hungary (2). The main partners in the UK are China (7), France (3), Brazil (2), Singapore (2), and the United States (2).

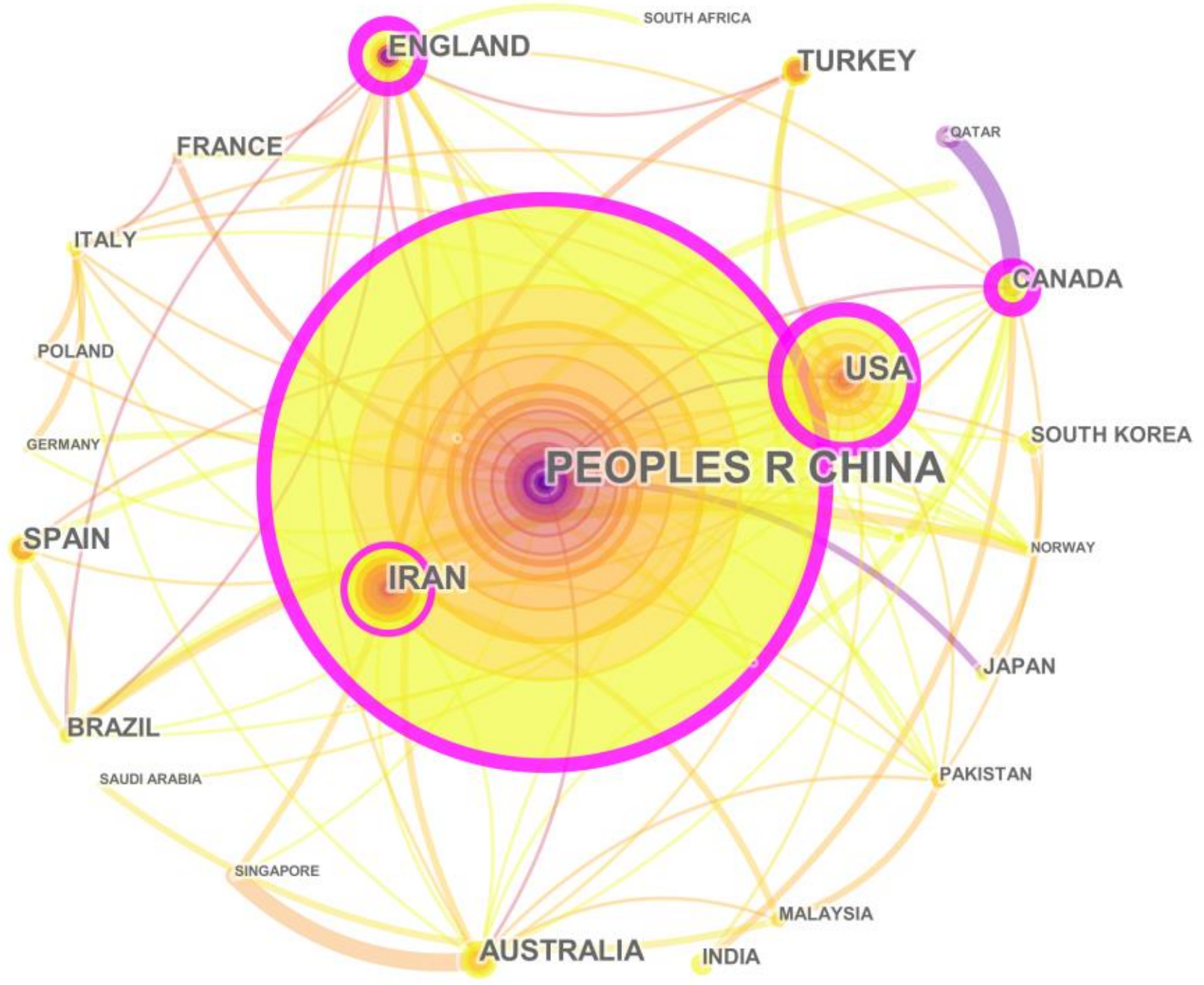

Fig.8. Country/region cooperation network in electricity demand forecasting.

# 5 Active topics and emerging trends

## 5.1 Co-occurrence network

Keywords are a clear sign of understanding the key content of research papers. Co-occurrence analysis is the number of times the term "statistics" appears in an article to measure the relationship between different articles and to further understand the research status in this field. The burst detection of keywords is often used for the emergence of hotspots and active topics in the field. Figure 9 shows a keyword co-occurrence network for power demand forecasting. For ease of observation, Figure 9 only retains points where the co-occurrence frequency is greater than 10. The key words co-occurrence network is intricate and complex, and the nodes are closely related, and mainly present the nouns and methods used in the research direction of the field. The keywords with a co-occurrence frequency higher than 100 are neural network (351), model (190), forecasting (170), demand (127), system (125), algorithm (123), and time series (109). We can find that the left part of the figure mostly refers to the main terms related to power demand forecasting such as forecasting, load, demand, consumption, etc., and the main methods involved in power demand forecasting are neural Network, model, algorithm, support vector regression, regression, etc.

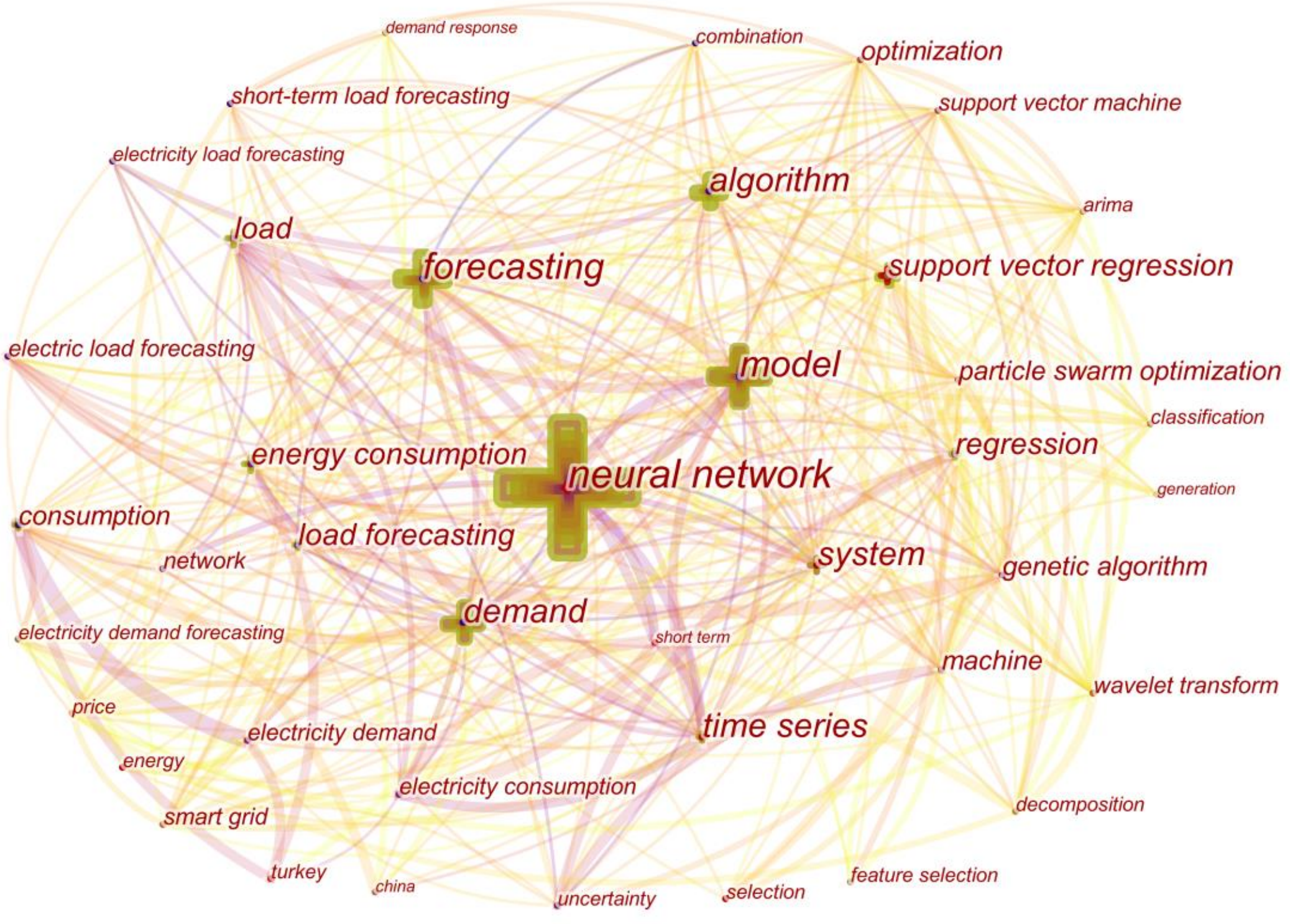

Fig.9. Keyword co-occurrence network in electricity demand forecasting.

Table 6 shows the 17 keywords with the highest burst detection. An entity with a frequency burst means that it has a sudden frequency change within a certain period of time. We found that the neural network is the most powerful keyword of burst and its duration is as long as 9 years (1999-2007), which indicates that the neural network is one of the most important basic methods in the field. It should be noted that the length of the whole line in the last columns of Table 8 represents the total research period (1991–2015) and the red line means the certain time period of the citation burst. At the same time, we also found that the larger keywords detected by burst are mostly method categories. This aspect is because the field of interest in this research is too small and the problems are concentrated. On the other hand, the new hotspots in this field are mostly methodological changes.

Table 6 Top 17 keywords with bursts during 1999-2018.

| **References** | **Year** | **Strength** | **Begin** | **End** | **1999 - 2018** |
|---|---|---|---|---|---|
| neural network | 1999 | 9.4035 | 1999 | 2007 | |
| system | 1999 | 4.277 | 1999 | 2003 | |
| implementation | 1999 | 4.9452 | 2002 | 2012 | |
| short term | 1999 | 4.8666 | 2005 | 2011 | |
| load forecasting | 1999 | 5.5095 | 2005 | 2006 | |
| time-series | 1999 | 7.3654 | 2006 | 2011 | |
| turkey | 1999 | 4.9077 | 2009 | 2011 | |
| electricity demand | 1999 | 3.3649 | 2010 | 2012 | |
| short-term load forecasting | 1999 | 4.4801 | 2012 | 2014 | |
| particle swarm optimization | 1999 | 4.2798 | 2013 | 2016 | |
| combination | 1999 | 3.8789 | 2013 | 2016 | |
| network | 1999 | 3.8621 | 2014 | 2016 | |
| intelligence | 1999 | 3.6923 | 2014 | 2016 | |
| selection | 1999 | 5.4265 | 2015 | 2018 | |
| energy | 1999 | 3.0846 | 2016 | 2018 | |
| support vector regression | 1999 | 3.7342 | 2016 | 2018 | |
| wavelet transform | 1999 | 3.4861 | 2016 | 2018 | |

The timeline in CiteSpace visualizes the cluster along the horizontal timeline (Figure 10). Each cluster is displayed from left to right. The legend at the top of the view shows the release time. Clusters are arranged vertically in descending order of their size. The largest cluster is shown at the top of the view. Color curves indicate co-occurrences in the year of the corresponding color. Large nodes or nodes with red tree rings are of particular interest because they are either co-occurring, or burst, or both. Below each timeline, the three most common references in a given year are displayed.

From Fig. 10, we can find that there are five clusters in the field of power demand forecasting, namely, keyword clustering, electric load forecast, regression analysis, neural network, daily peak load, non-linear autoregressive with exogenous inputs network and asymmetric error penalty function. The electricity load forecasting is mostly the research content, which runs through 20 years. Regression analysis and neural networks, nonlinear autoregressive and asymmetric penalty errors are key words involved in research methods in this field.

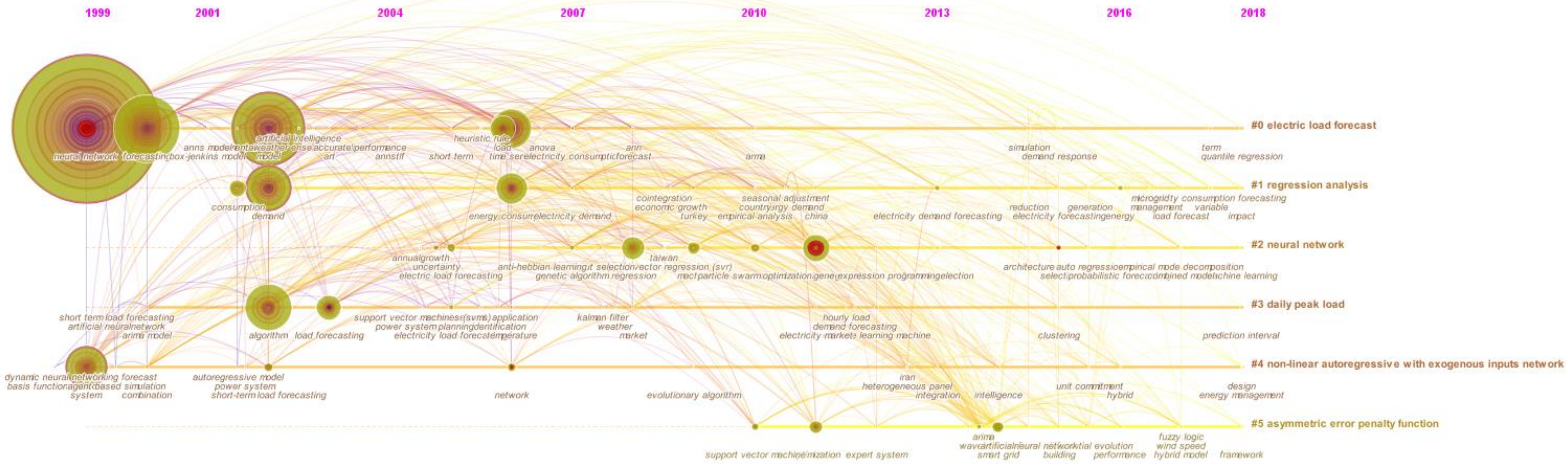

Fig.10. A timeline visualization for main keywords cluster

### 5.2 Co-citation network analysis

Co-citation network analysis is an analysis tool, usually used to examine a large number of documents and reveal the knowledge map of a scientific discipline. It analyzes and examines the related literature (such as documents, journals or authors) and is cited by other literature.

Table 7 Top 10 references with the strongest citation bursts during 1999-2018.

| References | Year | Strength | Begin | End | 1999-2018 |
|---|---|---|---|---|---|
| Bakirtzis et al. [36] | 1996 | 11.1269 | 1999 | 2006 | |
| Ramanathan et al. [37] | 1997 | 7.1414 | 2000 | 2007 | |
| Hippert [6] | 2001 | 20.7847 | 2002 | 2011 | |
| Darbellay and Slama[38] | 2000 | 8.8169 | 2003 | 2010 | |
| Pai and Hong [39] | 2005 | 6.1568 | 2006 | 2013 | |
| Taylor [22] | 2003 | 5.2395 | 2006 | 2013 | |
| Huang and Shih [40] | 2003 | 9.3716 | 2006 | 2013 | |
| Cottet and Smith [41] | 2003 | 4.4897 | 2006 | 2013 | |
| Amjady [42] | 2007 | 3.9629 | 2008 | 2015 | |
| Fan and Chen [43] | 2006 | 6.8201 | 2008 | 2015 | |

Table 7 lists the top 10 references with the strongest citation bursts. An article with citation burst means it received particular attentions from the academic circles in a certain period of time. The top-ranked item by bursts was Hippert (2001) with bursts strength of 20.7847. The second one was Bakirtzis (1996) with bursts strength of 11.1269.

By generating and analyzing the literature co-citation network in the power demand forecasting, the journal co-citation network and the author co-citation network can obtain the scientific knowledge structure and research frontier and history in this field.1 Literature has been cited. From the co-citation of the literature, we can find the development history and research frontier in the field of power demand forecasting. There are 875 nodes and 4429 edges in Figure 11.

From the chart below, we can find that the crucial articles in this field are mainly concentrated after 2009, which also shows that the articles in this field have experienced explosive growth in the next 10 years. The authors of important articles overlap with a large number of high-volume and high-cited authors in the field. Such as hong tao, fan s, taylor jw and so on. HIPPERT HS reviewed papers published between 1991 and 1999 to assess the application of neural networks in short-term load forecasting. Critically evaluate the design and testing of the neural networks presented in these papers. So it becomes a key node in the network [6]. Taylor JW compares the accuracy of short-term power demand forecasting with six univariate methods [28]. A semiparametric model is proposed to estimate the relationship between demand and model variables by Fan S [44]. An electric load forecasting model combining seasonal recursive support vector regression model and chaotic artificial bee colony algorithm (Srsvrcabc) is proposed by Hong WC [45]. Suganthi L review various energy demand forecasting models, including time series, regression, econometrics, ARIMA, and other traditional methods as well as such as fuzzy logic, genetic algorithms and neural networks, support vector regression, ant colony and particle swarm optimization [46].

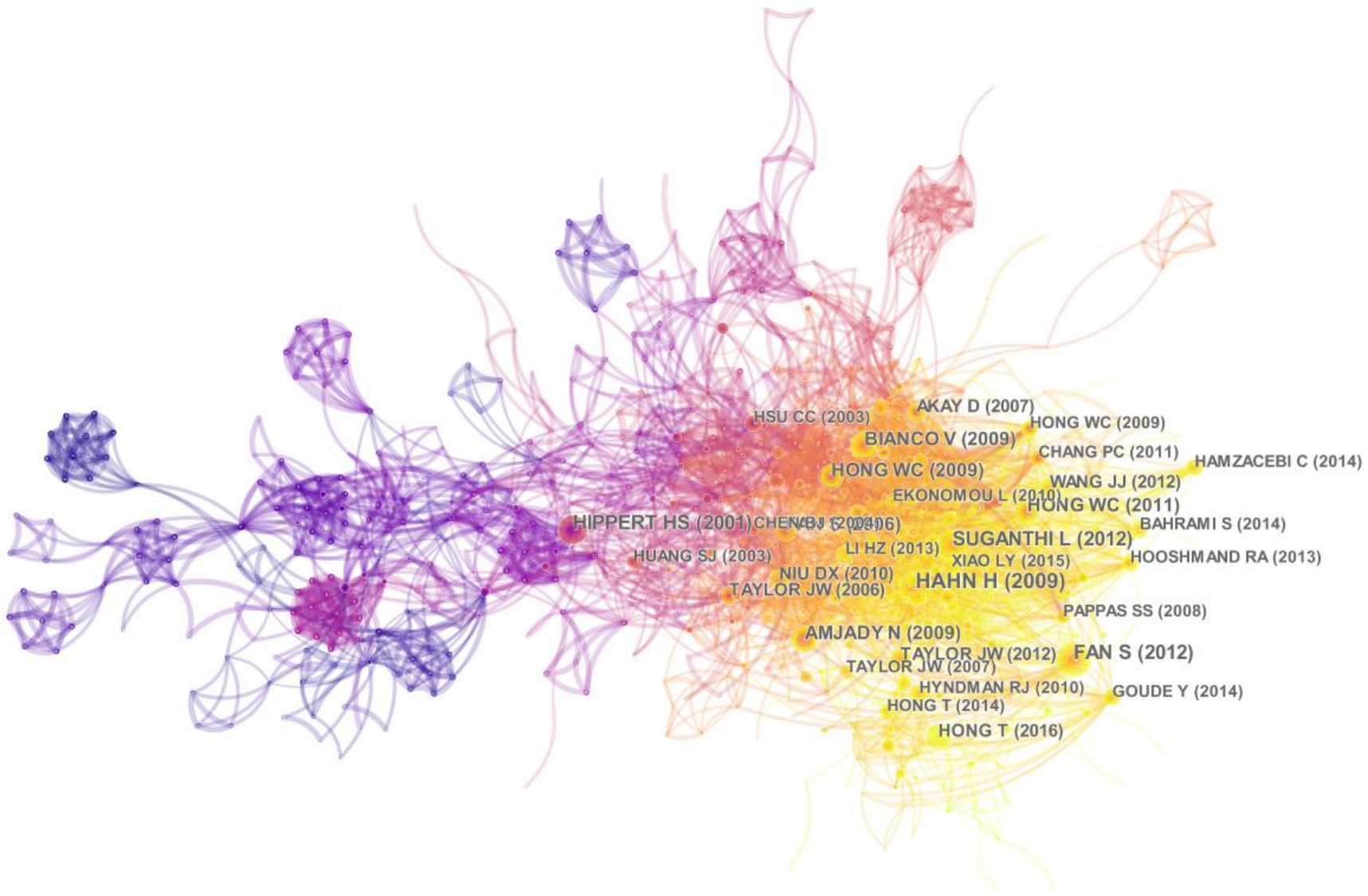

Fig.11. Reference co-citation network in electricity demand forecasting.

Fig.12 is a clustering network co-citation in the literature. It is obvious that the network has six clusters of a combined model, price forecasting, electricity consumption forecasting, peak load forecasting, support vector regression, neural networks, probabilistic forecasting, and turkey. Figure 13 is a line graph of the co-citation of the literature, similar to the keyword timeline. It shows the time evolution process of the six clusters. From the clustering results, we can find that the earliest clustering is neural networks, which is also the same as the keyword clustering results, which together illustrate the importance of neural networks in this field. The combined model is the largest cluster, and the time is close to 2018, indicating that the research frontier is a hybrid model. Price forecasting, electricity consumption forecasting, peak load forecasting and probabilistic forecasting reflect the main content of the research field. Price forecasting and peak load forecasting are the contents of earlier attention. Probabilistic forecasting and electricity consumption forecasting are more concerned in the near future. Support vector regression shows that it is one of the important methods in the field. Turkey was the main cluster between 1998 and 2008, indicating that during this period turkey's power forecasting was an area of concern. From the clustering results, we can find changes in the method of the field and changes in the content of the research. In particular, it is pointed out that the recent research method hotspot is the combined model.

In Table 8, Size indicates the numbers of the publications in the cluster. For example, the largest cluster (#0) has 140 members. Silhouette is an index to measure the homogeneity of a cluster, the greater value of this index, the better of the homogeneity. Mean (Year) represents the average year of the published documents of the regarding cluster. It is a very useful index since it can be used to judge the cluster whether is new or old.

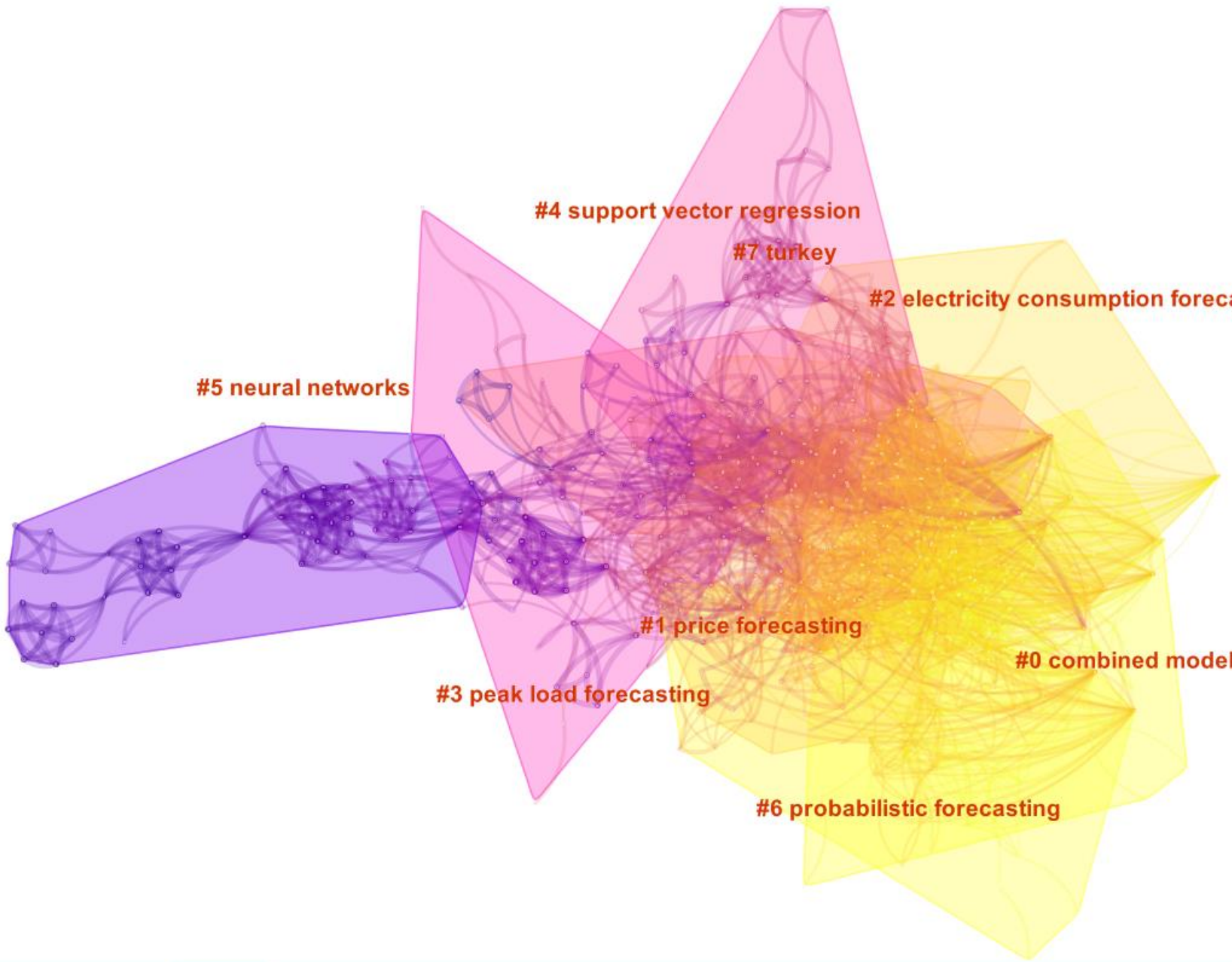

Fig.12. Main co-cited references cluster in electricity demand forecasting.

Table 8 Largest clusters of co-cited references, 1999-2018.

| Cluster ID | Size | Silhouette | Mean(Year) | Label (LLR) |
|---|---|---|---|---|
| 0 | 140 | 0.995 | 2013 | combined model |
| 1 | 98 | 0.996 | 2008 | price forecasting |
| 2 | 90 | 0.995 | 2010 | electricity consumption forecasting |
| 3 | 72 | 0.997 | 2001 | peak load forecasting |
| 4 | 58 | 0.998 | 2004 | support vector regression (svr) |
| 5 | 57 | 0.998 | 1995 | neural networks |
| 6 | 56 | 0.998 | 2011 | probabilistic forecasting |
| 7 | 53 | 0.999 | 2003 | turkey |

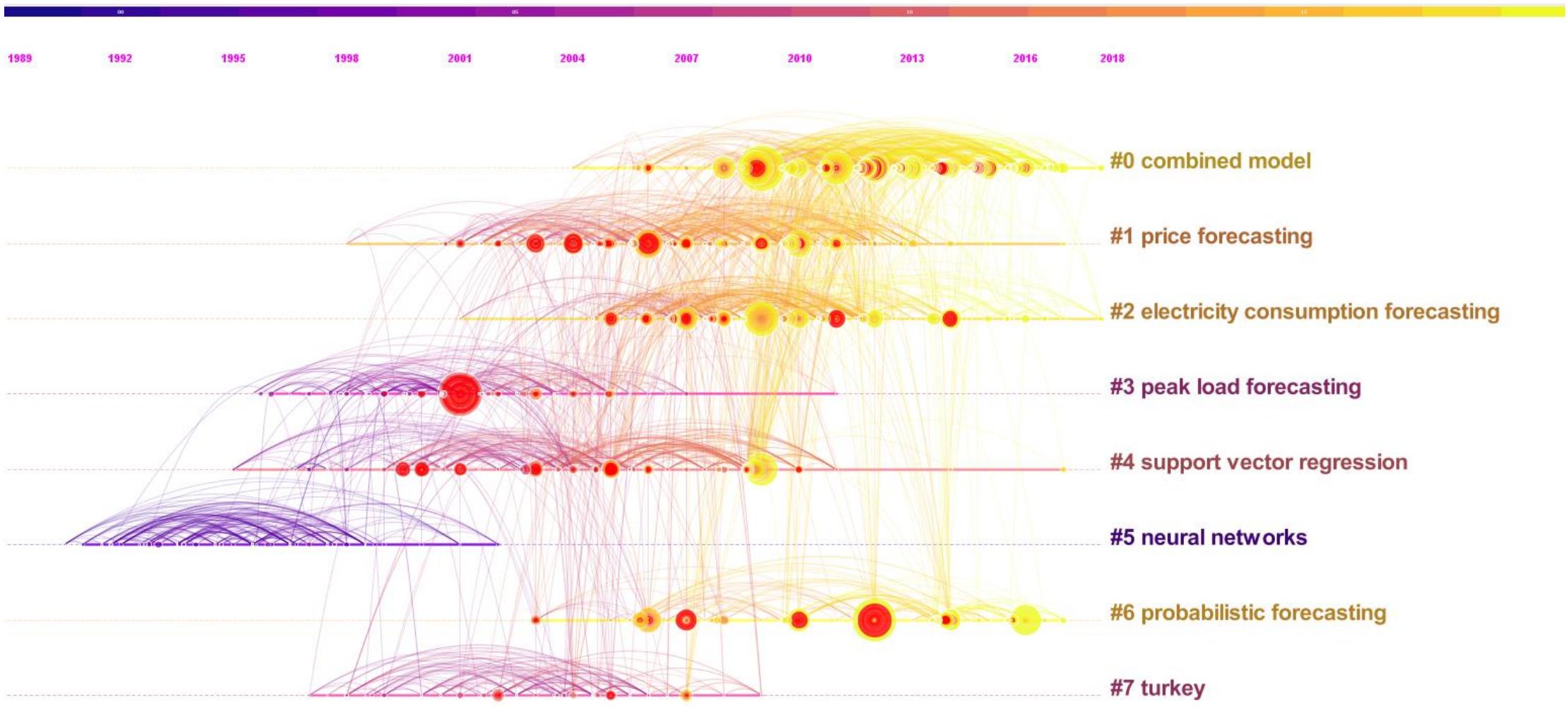

Fig.13. A timeline visualization for main references cluster

## 6 Conclusions

In this paper, we made a scientometric review and visualization analysis on electricity load forecasting studies based on 831 publications retrieved from Web of Sciences. An integrated knowledge map of the electricity load forecasting field and hot topics with emerging trends are presented by collaboration network, keywords co-occurrence analysis and co-citation analysis. Some interesting and useful conclusions are as follows.

First, electricity load forecasting has received more and more attention, the numbers of citations and publications are increasing rapidly, especially in the last decade. “Energy fuels”, which accounts for 36.82%, is the largest subject category in the electricity load forecasting research area. “Energy” is the highest yield journal with 81 publications, followed by “Energies”, “International Journal of Electrical Power Energy Systems” and “Applied Energy”. “Renewable & Sustainable Energy Reviews”, “European Journal of Operational Research” and “Applied Energy”, are of constant interest to researchers in the field recently. North China Electric Power University, Lanzhou University, University of North Carolina, Islamic Azad University and Oriental Inst Technol is the top 5 yield Institution. Wang JZ, which publishing 24 articles in the field, is the most high-yield author, followed by Hong WC, Niu DX and Hong T. The publications of Zareipourh, Khosravia, Hong T, Abediniao and Guo S have received a lot of attention in recent years, and the publications of Fan S have received a lot of attention all the time.

Second, there are 2143 scholars involved in the field of power demand forecasting, but most of them only cooperate in a very small scope and almost cooperate with authors in the same institution. The largest cooperative networks were formed with Wang Jianzhou who is the largest Structural hole in the cooperative networks. North China Electric Power University, Lanzhou University, Islamic Azad University, Oriental Inst Technol, and University of Tehran are the five most irreplaceable contributors and productive institutions in the field of electricity load forecasting. Peoples Republic of China (including Taiwan, Hong Kong and Macau), USA, Iran, England and Turkey are the 5 largest contributors to national and regional cooperation networks.

Third, neural network, support vector regression and combined model are main methods in electricity load forecasting, in addition, support vector regression, combined model and wavelet transform are hotspots methods. price forecasting, electricity consumption forecasting, peak load forecasting and probabilistic forecasting are main researches in electricity load forecasting, especially, probabilistic forecasting and electricity consumption forecasting are hotspots.

The basic situation of subject classification, journals, authors, institutions, countries and highly cited papers in the electricity load forecasting can be figured out, based on the research of this paper. At the same time, the corresponding collaboration of countries/regions, institutions and authors are also studied in our research. Furthermore, emerging trends and new developments in this area are discussed in this research. The results of this research provide a comprehensive description of the electricity load forecasting and are helpful for scholars to maintain the development of this field.

The limitations in our research are that, due to the limits of co-citation analysis in citespace, the literature in our paper are only retrieved from the core database of WoS, Document types are limit in “article” or “review” and Literature type are limit in “English”, which may make some significant literature have been overlooked.

## Conflict of interests

The authors declare that there are no conflicts of interest regarding the publication of this paper.

## Acknowledgments

This research work was partly supported by the National Natural Science Foundation of China under Grants No. 71774130 and No. 71988101. supported by the Fundamental Research Funds for the Central Universities under Grant No. xpt012020022.